\documentclass{iaus}
\usepackage{graphicx}

\title[short title of paper] 
{The Millennium Galaxy Catalogue: Galaxy Bimodality}

\author[short author list]   
{Simon P. Driver$^1$, Jochen Liske$^2$ \break \and Alister W. Graham$^3$}

\affiliation{$^1$ School of Physics and Astronomy, University of St Andrews, St Andrews, Scotland,
\break email: spd3@st-and.ac.uk\\[\affilskip] 
$^2$ European Southern Observatory, Karl-Schwarzschild-Str. 2, 85748 Garching, Germany, \\[\affilskip] 
$^3$ Astrophysics and Supercomputing, Swinburne University of Technology, Australia}

\pubyear{2006}
\volume{235}  
\pagerange{1--2}
\date{?? and in revised form ??}
\jname{Galaxy Evolution across the Hubble Time}
\editors{F.Combes \& J. Palous, eds.}
\begin{document}

\maketitle

\begin{abstract}
Galaxy bimodality is caused by the bulge-disc nature of galaxies as
opposed to two distinct galaxy classes. This is evident in the
colour-structure plane which clearly shows that elliptical galaxies
(bulge-only) lie in the red compact peak and late-type spiral galaxies
(disc-dominated) lie in the blue diffuse peak. Early-type spirals
(bulge plus disc systems) sprawl across both peaks. However after
bulge-disc decomposition the bulges of early-type spirals lie
exclusively in the red compact peak and their discs in the blue
diffuse peak (exceptions exist but are rare, e.g., dust reddened
edge-on discs and blue pseudo-bulges). Movement between these two
peaks is not trivial because whilst switching off star-formation can
transform colours from blue to red, modifying the orbits of $\sim 1$
billion stars from a planar diffuse structure to a triaxial compact
structure is problematic (essentially requiring an equal mass
merger). We propose that the most plausible explanation for the dual
structure of galaxies is that galaxy formation proceeds in two stages.
First an initial collapse phase (forming a centrally concentrated core
and black hole), followed by splashback, infall and accretion (forming
a planar rotating disc). Dwarf systems coule perhaps follow the same
scenario but the lack of low luminosity bulge-disc systems would imply
that the two components must rapidly blend to form a single flattened
spheroidal system.  
\end{abstract}

\firstsection 
\section{Introduction}
\begin{figure}
\includegraphics[height=4.0in,width=5.5in]{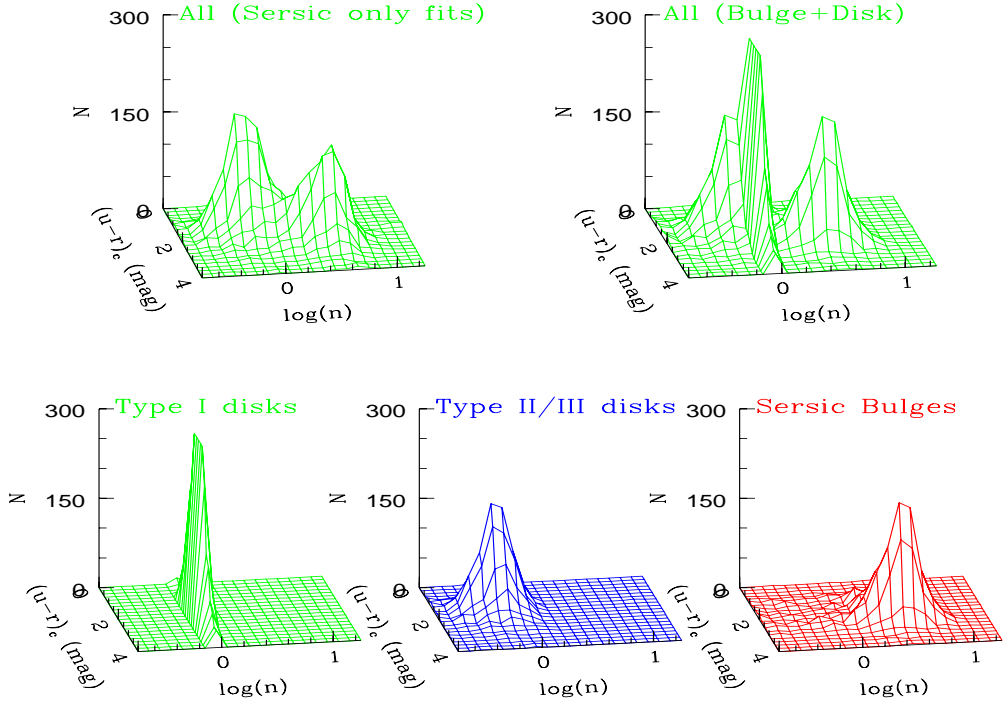}
\caption{Galaxy bimodality in the colour-structure plane, see text for details.}\label{driver_fig}
\end{figure}

The Millennium Galaxy Catalogue (\cite{mgc01}; \cite{mgc03}) contains
10095 galaxies to $B<20$ mag extending over a 37 deg$^2$ region of
the equatorial sky. The redshift survey is 96\% complete
(\cite{mgc04}) and most galaxies have sufficient resolution to achieve
reliable bulge-disc decomposition (\cite{mgc08}). We have explored
galaxy bimodality in the colour-structure plane using our sample and
find an extremely strong bimodal distribution in this plane (see
Fig.~\ref{driver_fig} and also \cite{mgc05}). For colour we use the
PSF $(u-g)_c$ from SDSS DR1 (\cite{sdssdr1}) and to represent
structure we use the S\'ersic index (\cite{gd05}).
Fig.~\ref{driver_fig} upper left shows the observed bimodal
distribution for all galaxies brighter than $M_B -5\log h = -17$ mag
based on the {\it global} S\'ersic index. The upper right panel shows
the equivalent plot but with the colour and S\'ersic index now derived
for the {\it distinct} bulge and disc components.  In the lower panels
we show the dominant stellar structures for clarity, which together
define the nearby galaxy population. These are: exponential discs
($n=1$), truncated and anti-truncated discs ($n<1.2$), and bulges ($n
> 1.2$).The lack of significant overlap between the two disc populations and the
bulge populations lead us to infer distinct formation mechanisms for
bulge and disc systems.  The coincidence of the two disk populations
implies a single common disc formation process and inside out growth.

\section{Conclusions}\label{sec:concl}
\noindent
{\bf 1.} Galaxies are comprised of two fundamental stellar
   configurations: bulges and discs.

\noindent
{\bf 2.} Global measurements such as the star-formation rate and size
   evolution may be misleading (non-optimal) approaches and one
   should strive for distinct bulge and disc measurements instead.

\noindent
{\bf 3.} Luminous galaxies have {\it probably} formed in two stages,
   early collapse to form a triaxial pressure supported bulge (with
   black hole), followed by splashback, infall and accretion to form a
   planar rotating disc.

\noindent
{\bf 4.} Truncation of galaxy discs is common and implies ongoing
   inside-out growth.

\noindent
{\bf 5.} {\it If} dwarf systems adhere to this scenario it may be that
   their bulge and disc components are unstable and blend. Some
   evidence for this is perhaps seen in the DART survey of the Fornax dwarf
   spheroidal (see \cite{batt06}).

\noindent
More details on galaxy bimodality are presented in Driver et al. (2006).

\end{document}